\documentclass[acmsmall,screen,nonacm]{acmart}

\usepackage{listings}
\usepackage{todonotes}
\usepackage{colortbl}
\usepackage{wrapfig}
\setcopyright{none}
\renewcommand\footnotetextcopyrightpermission[1]{}

\definecolor{diffminus}{RGB}{176,0,32}
\definecolor{diffplus}{RGB}{0,112,0}
\definecolor{promptadded}{RGB}{190,245,190}
\definecolor{promptremoved}{RGB}{255,190,190}

\begin{document}
\raggedbottom
\setlength{\emergencystretch}{1em}

\title{Distilling Reasoning Traces into Advisory Prompts for Software Engineering Tasks}

\author{Faizan Faisal}
\affiliation{%
  \institution{University of California, Davis}
  \city{Davis}
  \state{California}
  \country{USA}
}
\email{fznfaisal@ucdavis.edu}

\author{Prem Devanbu}
\affiliation{%
  \institution{University of California, Davis}
  \city{Davis}
  \state{California}
  \country{USA}
}
\email{ptdevanbu@ucdavis.edu}

\author{Toufique Ahmed}
\affiliation{%
  \institution{IBM}
  \country{USA}
}
\email{tfahmed@ibm.com}

\begin{abstract}
Language models are widely used for generating and otherwise processing code 
(e.g., predicting behaviour, 
outputs, or likely inputs) 
and this paper asks whether better performance can be achieved without having to always incur the cost of reasoning.
Human students of programming learn to avoid mistakes by a) first identifying them, b) reflecting
upon cognitive lapses that led to these mistakes (essentially, ``thinking through'' the errors), c) inferring general rules/lessons from these ``thinking through'' reflections, and
d) internalizing these lessons into rules. In tutorial sessions with an instructor, this is a
very common Socratic interaction with students. Examples of such internalizable rules
might include the nugget \emph{``Before coding, restate the requirements to clarify them''}.
Inspired by this process, this paper describes an approach where
we first identify examples in which ``thinking mode'' in a (low-resource) LLM avoids errors; then, these
errors, and their avoidance via ``thinking'' in the same LLM, are examined by a bigger LLM to generate summary explanations; these are then summarized by a large LLM into \emph{brief} advisory prompts. This approach
works on many modestly sized models; in some cases, the ``advisory prompts''
thus learned can also be gainfully transferred to other models. We also present some investigations into the nature of coding errors that language models make, and a characterization of when this
approach can be helpful.

\end{abstract}



\maketitle

\section{Introduction}
\label{sec:introduction}

\begin{figure}[tb]
\centering
\begin{minipage}{0.7\linewidth}
{\footnotesize\noindent\textbf{(a) Stage 1: Source delta-case diagnosis.}\par}
\fbox{\parbox{0.94\columnwidth}{\footnotesize\itshape
\emph{\textbf{Mechanism explanation:}} Reasoning improved the answer by recognizing that evaluating \texttt{array[i] < 0} on the string \texttt{'aaa'} raises a \texttt{TypeError}...

\emph{\textbf{Prompt suggestion:}} Check each operation for type compatibility before deciding whether execution succeeds.
}}

{\footnotesize\noindent\textbf{(b) Stages 2--3: Distilled prompt construction.}\par}
\fbox{\parbox{0.94\columnwidth}{\footnotesize\itshape
Trace execution in order on the concrete inputs, \textcolor{diffplus}{\textbf{checking runtime types}}, API preconditions, mutations, and loop/recursion state to identify \textcolor{diffplus}{\textbf{the first operation that can fail}} or not terminate...}}

{\footnotesize\noindent\textbf{(c) Stage 4: Held-out evaluation.} For input \texttt{['']}, the comparison \texttt{'' < 0} itself raises \texttt{TypeError}.\par}
\begin{lstlisting}[language=Python,basicstyle=\footnotesize\ttfamily,numbers=none,xleftmargin=0pt,frame=single,aboveskip=1pt,belowskip=3pt,escapechar=@]
def f(array):
    for i in range(len(array)):
        if array[i] < 0:  # "" < 0 raises TypeError
            array.pop(i)
    return array

# input: ['']; expected: TypeError
@\textcolor{diffminus}{\textbf{- baseline: None}}@
@\textcolor{diffplus}{\textbf{+ distilled: TypeError}}@
\end{lstlisting}
\end{minipage}

\caption{Exception-prediction example: Stage 1 diagnoses a type-compatibility error, Stages 2--3 abstract it into a distilled prompt, and Stage 4 shows the held-out correction.}
\Description{Three-part example of the prompt-distillation process. A teacher
diagnoses a runtime type error, distills the diagnosis into a general
type-compatibility instruction, and the instructed student correctly predicts
a TypeError on a held-out Python example.}
\label{fig:motivating-example}
\end{figure}

LLMs have been widely used for many software engineering tasks,
such as code generation~\citep{jain2025livecodebench}\footnote{We include only just one illustrative citation, for
brevity; all these areas are quite large and
expanding.}, GitHub issue resolution~\citep{jimenez2024swe}, input/output prediction~\citep{gu2024cruxeval}, etc. 
However, LLMs do make mistakes. Whether these mistakes arise from bugs in training 
data~\citep{al2026model} or from mistaken generalization~\citep{berglund2024reversal}, 
once such mistakes are identified, it would be desirable to find systematic ways to avoid repeating them. An error-free
training corpus might help; however,
in the case of software,
defects are nearly impossible to avoid, even in released code; so a bug-free training corpus would be very difficult to collect; 
even if we could, retraining is very costly. Therefore, training-free, inference-time methods,
especially prompt engineering, would be attractive options. Such training-free error reduction is especially important for small language models (SLMs), which are inexpensive, fast, and deployable on resource-constrained devices~\citep{lu2024small,nguyen2025survey}; improving their reliability enables them to become practical alternatives to much larger models. But how can this be done? We take inspiration from human learning. 

Beginning (human) programmers also  make mistakes; but over time, they typically learn to avoid errors by reflecting upon, and explaining/understanding their past coding mishaps (perhaps with instructor assistance). These explanations gradually help students extract and internalize \emph{general advice} that can be considered in future development tasks. Inspired by this human learning process, we seek an automated procedure
that can  mechanically distill general ``advisory prompts'' that, when
injected into a task prompt, enhance LLMs' behaviour,
and help them sidestep known past errors on tasks. For example, Figure~\ref{fig:motivating-example} shows one such example, for the task
of predicting outputs, where the distilled advisory prompt helps the model work correctly. The error in
this case concerns
runtime type compatibility; our approach distills 
error diagnosis into an advisory prompt that helps the same model correctly recognize a \texttt{TypeError} on many held-out cases. 
On this example, the distilled prompt,
(which uses a non-thinking response) produces
correct output; in addition, it
emits just 246 output tokens, compared with 312 for the correct thinking-mode baseline: 66 fewer tokens, or a 21.2\% saving.

Our approach is inspired by the hypothesized student learning iteration above; typically,
we improve  the performance of a small, hybrid reasoning model $M$.  
We first run this
model $M$ in non-thinking and thinking (\emph{``reflective''}) modes, and identify the set $\mathcal{C}$ of distillation cases for a given task where
thinking flips the performance (wrong to right, or \emph{vice versa}). For each case $i \in \mathcal{C}$, assume that $y^*_i$ is the right answer. Now, when $M$ runs on case $i$, it gives us a non-thinking answer $\hat{y}_{i,0}$, and
a thinking answer $\hat{y}_{i,1}$, together with its reasoning (\emph{``explanation''}) trace $T_{i,1}$.
We then present $\langle \hat{y}_{i,0},\hat{y}_{i,1}\rangle$ to a frontier ``teacher'' model with the reasoning trace $T_{i,1}$  and the \emph{correct} answer $y^*_i$, then ask 
the frontier model to process
why the reasoning helped, or hurt, and then
to summarize the behavior as a reusable prompt suggestion $P_i$. We then present
the entire collection of suggestions $\{P_i, i \in \mathcal{C}\}$ to the teacher, and ask it to \underline{\emph{distill}} a candidate advisory prompt template $P^*$ that would likely help $M$
(and perhaps even other models) avoid
most of the errors in $\mathcal{C}$, \emph{even without reasoning}. Thus, 
our goal here is to 
improve correctness, \emph{while also} incurring
lower token expenditure than when using full reasoning.


We would certainly expect the selected advisory prompt template $P^\star$ thus derived to help
model $M$ to largely avoid the errors in $\mathcal{C}$;
however, we would like to know if $P^\star$ would help avoid errors in held-out sets
that were not used to generate it.

\textbf{RQ1:} Can advisory prompts distilled from thinking/non-thinking correctness deltas improve held-out performance on reasoning-intensive software-engineering tasks?

We note that the selected template $P^\star$ is derived by summarizing the right and wrong answers (and full reasoning traces) from $M$ in cases where thinking by $M$ changed the answer. Thus,
we might expect that $P^\star$ contains remedial
instructions that digest the reasoning tokens; thus using $P^\star$ with $M$ should improve accuracy while expending fewer tokens.

\textbf{RQ2:} Can the distilled advisory prompt provide improved accuracy even
in the non-thinking case, while expending fewer output tokens than in the full thinking case?

Recent work suggests that programs generated by different language models from the same specification (just like programs written by different humans~\citep{knight1986experimental}
from the same spec)
tend to fail unexpectedly often on the same inputs, \emph{viz.}, they
have \emph{common-mode} failures~\citep{ron2026version}; this suggests
that language models tend to fail  in similar ways. This led us to ask if
an advisory prompt distilled from one model's errors could help another model (whose errors
may be similar).

\textbf{RQ3:} Can an advisory prompt distilled from errors made by one model $A$ help another LLM $B$ avoid errors on the same reasoning-intensive software-engineering task?

We also examined whether the avoided errors were shared.

\textbf{RQ4:} When an advisory prompt from one model $A$ helps improve performance on a different model, $B$, is it because both $A$ and $B$ fail on the same examples?

We experiment with four software-engineering task families---code generation on LiveCodeBench, program input prediction, output prediction, and exception prediction---using four modestly sized hybrid reasoning models as students.

Alongside the four research questions above, we report controlled studies
that isolate the information in the reasoning trace, compare the
pipeline with the GEPA prompt optimizer under matched folds and decoding, and
measure prompt growth, teacher-side optimization cost, and deployment-cost scaling.

\section{Methodology}
\label{sec:methodology}

\subsection{Problem Formulation}

Let $\mathcal{D}=\{(x_i,y_i,m_i)\}_{i=1}^n$ be a benchmark dataset, where $x_i$ is the task input, $y_i$ is the reference answer or executable oracle, and $m_i$ contains metadata such as release, difficulty, or task source. Each task provides a scoring function $s(\hat{y}_i,y_i)\in\{0,1\}$ indicating whether model response $\hat{y}_i$ passes the task-specific evaluation: pass/fail under the benchmark tests for code generation, and exact or task-specific correctness for the other software-engineering tasks. A prompt template $\pi$ maps an input $x_i$ to the rendered prompt $\pi(x_i)$. The student model is a hybrid reasoning model \citep{qwen2025qwen3}: with thinking mode enabled it emits a full reasoning trace before its final answer, and in non-thinking mode it answers directly without a full reasoning trace. Throughout, \emph{reasoning trace} refers to the intermediate tokens generated between the model's think delimiters (e.g., \texttt{<think>} and \texttt{</think>}), separate from the final response. For student model $M$, baseline template $\pi_0$ (which applied to each example $x_i$), and mode indicator $z\in\{0,1\}$ ($0$: non-thinking; $1$: thinking), we record $\hat{y}_{i,z}=M(\pi_0(x_i),z)$ and $a_{i,z}=s(\hat{y}_{i,z},y_i)$; the paired behavioral delta is $d_i=a_{i,1}-a_{i,0}$. Cases with $d_i=1$ are \emph{improvement cases} (the thinking-mode sample succeeds where the non-thinking sample fails); cases with $d_i=-1$ are \emph{regression cases} (the reverse). We refer to these paired changes as thinking/non-thinking correctness deltas: the delta identifies which examples are informative, while the paired responses and reasoning trace provide evidence about procedural differences to embrace or avoid. The paired design holds the example, prompt template, expected answer, model, and evaluation oracle fixed.

\subsection{Prompt Distillation Pipeline}

We study whether thinking mode's performance gains can be distilled into summary prompt instructions, yielding an augmented prompt we call a \emph{distilled prompt} (\emph{aka} ``advisory prompt'')\footnote{We use \emph{prompt distillation} to mean synthesizing natural-language instructions from model behavior. This differs from distilling prompted knowledge or context into model weights \citep{snell2022learning,kujanpaa2025knowledge} or compressing discrete prompts into soft prompt vectors \citep{li2023pod}; we do neither.}. We operationalize the approach sketched in the introduction; the teacher uses two frontier models (Section~\ref{sec:experimental_setup}). The procedure is inference-only, with no weight updates. After paired baseline measurement and correctness-delta mining, it consists of four stages: case-level diagnosis, instruction aggregation, candidate generation and validation selection, and held-out evaluation.

\begin{figure}[tb]
\centering
\includegraphics[width=0.7\linewidth,draft=false]{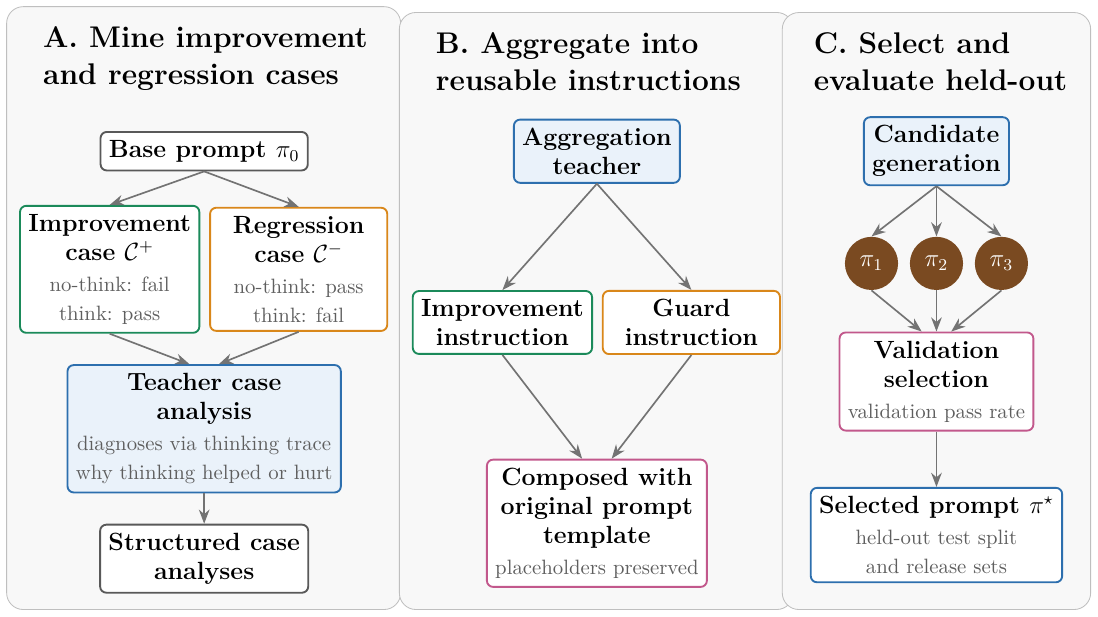}
\caption{The prompt-distillation pipeline: correctness-delta mining, teacher diagnosis of improvement and regression cases, aggregation into improvement and guard instructions, and validation-based selection before held-out evaluation.}
\Description{Flow diagram showing correctness-delta mining before four stages:
diagnose correctness-changing cases, aggregate diagnoses into reusable
instructions, generate and select candidate advisory prompts on validation
data, and evaluate the selected prompt on held-out data.}
\label{fig:prompt_distillation_pipeline}
\end{figure}

The pipeline has 4 stages, shown in Figure~\ref{fig:prompt_distillation_pipeline}. Before Stage~1, we run the student model, with the baseline prompt, on a split reserved for distillation. 
For this split, we run with \emph{and} without thinking; in both cases we store prompts, predictions, scores, response lengths, and reasoning traces. From the scored, paired 
(thinking/non-thinking) samples,  we form the 2 delta sets: the improvement set $\mathcal{C}^{+}=\{i:a_{i,0}=0,a_{i,1}=1\}$ and the regression set $\mathcal{C}^{-}=\{i:a_{i,0}=1,a_{i,1}=0\}$.

In Stage~1, a teacher model analyzes each delta case, including: the task prompt, expected answer or oracle summary, both student responses, scores, and the available reasoning trace 
(Note that this
is only the distillation split; the models don't see the
validation and test examples yet). The teacher model compares the two responses, and diagnoses why the trace appears to have helped (improvement cases) or hurt (regression cases). The teacher's output for each case is well-structured: each  analysis output includes a mechanism label (a brief behavior tag; e.g., ``checks boundary cases''), a short explanation, a prompt suggestion, an evidence span, and a confidence assessment (how confident the model is). For improvement cases, the suggestion describes what the student should do more systematically; for regression cases, it instructs against unsupported assumptions, unnecessary transformations, or changes to required output formats.

In Stage~2, the teacher aggregates the gathered set of analysis outputs
 into concise, reusable improvement instructions and guard instructions. Improvement instructions prescribe actions to be done more systematically, such as checking feasibility conditions, deriving invariants or validating boundary cases. Guard instructions discourage patterns associated with regressions. In Stage~3, these instructions are composed with the original task template to form a finite candidate set of
distilled prompt templates $\Pi=\{\pi_1,\ldots,\pi_k\}$. Candidate templates may differ in wording, ordering, and how strongly they combine improvement and guard instructions, but they must preserve the original task fields and placeholders.

Still in Stage~3, each candidate template is evaluated on the unseen \emph{validation split} under both inference modes, and we select the prompt template $\pi^\star=\arg\max_{\pi_j\in\Pi}\widehat{p}_{\mathrm{val}}(\pi_j)$, where $\widehat{p}_{\mathrm{val}}$ is the mean of the thinking and non-thinking validation pass rates. Ties are resolved, in order, by fewer baseline-success regressions (where the distilled prompt flips a pass to a fail), shorter prompt length, and lower candidate prompt index. The original template is not included as a candidate.  In Stage~4, the selected prompt is evaluated on the held-out test fold or release set.
Figure~\ref{fig:prompt-example} shows a representative baseline-to-distilled
prompt diff for Exception Prediction: the selected prompt replaces the
original instruction while preserving the task wrapper and input placeholders.

Finally, we note that Stages 1-4 are possible only if delta cases exist; otherwise, there is no signal to distill; if only one direction is present, candidates use that direction alone; neither occurred in our experiments.

\begin{figure}[tb]
\centering
\textbf{Baseline-to-distilled prompt diff}
\begin{lstlisting}[
  basicstyle=\scriptsize\ttfamily,
  numbers=none,
  frame=single,
  rulecolor=\color{black!55},
  backgroundcolor=\color{white},
  xleftmargin=0pt,
  breaklines=true,
  escapeinside={(*@}{@*)},
  aboveskip=3pt,
  belowskip=3pt
]
You are solving a Python runtime-failure prediction task. 

Given the Python function and its input arguments,
(*@\begingroup\setlength{\fboxsep}{2pt}\colorbox{promptremoved}{\parbox{\dimexpr\linewidth-2\fboxsep\relax}{\scriptsize\ttfamily predict whether execution will raise an exception. If an exception occurs, return the exact exception type (for example: IndexError, KeyError). If the code times out return TimeoutError. If no exception occurs, return None.}}\endgroup@*)
(*@\begingroup\setlength{\fboxsep}{2pt}\colorbox{promptadded}{\parbox{\dimexpr\linewidth-2\fboxsep\relax}{\scriptsize\ttfamily trace execution on the concrete runtime values step by step. Treat the given call literally: first verify argument arity, concrete runtime types, and the validity of the first operation, including method/API preconditions, bounds, and termination. Do not repair malformed input or speculate about later behavior; predict the earliest actual runtime failure. If execution raises an exception, return the exact exception type (for example: IndexError, KeyError). If the code times out, return TimeoutError. If no exception occurs, return None.}}\endgroup@*)

At the end of your answer, output one line in this format:
Answer: <exception type or TimeoutError or None>

Function:
```python
{code}
```

Input arguments:

```python
{input}
```
\end{lstlisting}
\caption{Prompt example: Baseline-to-distilled template
diff for Exception Prediction.  The red block is the instruction removed from
the baseline, the green block is its distilled replacement, and the remaining
text is shared by both templates.  The placeholders
\texttt{\{code\}} and \texttt{\{input\}} are filled for each instance.}
\Description{A baseline-to-distilled prompt diff for predicting Python runtime
exceptions.  A red block marks the removed baseline instruction and a green
block marks the distilled replacement containing step-by-step execution,
literal-call, arity, runtime-type, API-precondition, bounds, termination, and
earliest-failure checks.  The remainder of the prompt is shared.}
\label{fig:prompt-example}
\end{figure}

\section{Experimental Setup}
\label{sec:experimental_setup}

\subsection{Benchmarks}
\label{sec:benchmarks}
First, or code generation, we use LiveCodeBench \citep{jain2025livecodebench},
which provides six releases (v1--v6). We use the 342 examples in the newer
v5/v6 releases for prompt construction and five-fold held-out evaluation.
Within each fold, validation performance selects one prompt from three
candidates. This produces five selected prompts per model, one from each fold,
and 20 prompts across the four models.
We then evaluate on the older v1--v4 releases. To limit cost, for each model
and each older release, we randomly select one of the model's five prompts
learned from the v5/v6 folds. This avoids choosing a prompt based on its
performance on the older release. We apply the selected prompt to every
example in that release, giving 713 examples across v1--v4. The LiveCodeBench
result in Table~\ref{tab:rq1_effects} pools the v5/v6 held-out fold
evaluations with these four older-release evaluations.


We use three additional software-engineering datasets as secondary validation.
\emph{Input Prediction} is based on the CRUXEval input-prediction task
(CRUXEval-I) \citep{gu2024cruxeval} and contains 800 examples. The model must
infer an input consistent with a given program behavior or output.
\emph{Exception Prediction} contains 834 examples from the exception-prediction
extension of CRUXEval-style output prediction introduced by
\citet{spiess2026robustly}. The model is given a program fragment and must
predict whether its execution raises an exception and, if so, which one; this
is a runtime-exception prediction task in the spirit of ThrowBench
\citep{prenner2025throwbench}. Finally, \emph{Output Prediction} asks the model
to predict a program's output from its input. It contains 7,450 examples from
the perturbed-inputs variant of the CRUXEval-style output-prediction task
\citep{spiess2026robustly}. Programs are held fixed, while test inputs are
replaced by type-aware mutated inputs with execution-derived reference
outputs. We use this variant because models score near ceiling on the original
examples but drop sharply under input perturbation, making it a more reliable
measure of execution reasoning.


\subsection{Models}

We evaluate prompt distillation on four student models: Qwen3-4B and Qwen3-8B \citep{qwen2025qwen3}, and Gemma-4-E2B-it and Gemma-4-E4B-it \citep{gemmateam2026gemma}, abbreviated Gemma-4-E2B and Gemma-4-E4B in tables.  All four are hybrid reasoning models that expose both a thinking and a non-thinking inference mode; each student model is evaluated under both modes, using the same base task prompt. We use teachers only during prompt distillation. GPT-5.4-mini performs Stage~1 case analysis (Figure~\ref{fig:motivating-example}a), GPT-5.4 performs Stages~2--3 aggregation and prompt construction (Figure~\ref{fig:motivating-example}b), and the student models perform all Stage~4 held-out evaluations (Figure~\ref{fig:motivating-example}c).

\subsection{Main Protocol}

The primary analyses use five folds for each task family. For
LiveCodeBench, these folds cover v5/v6; v1--v4 are separate release
evaluations. In each fold, 20\% of the examples are held out for testing, and the remaining 80\% is split into 87.5\% distillation and 12.5\% validation, yielding approximately 70/10/20 overall. The distillation partition supplies the delta cases used to synthesize prompt instructions, the validation partition selects among candidate prompts, and the test partition is used only for held-out evaluation. Stage 3 generates $k=3$ candidate prompt templates, per fold. Student and teacher models use temperature 0.7. The reasoning effort for teacher models is set to none. Student models have a 14k-token output budget and generate one sample per example. For every model, dataset, and inference mode, the baseline and selected distilled prompt are compared on paired examples using the evaluation metrics defined below; because we draw a single sample per example, pass rate coincides with pass@1 \citep{chen2021evaluating}. We observed Qwen models to produce excessively long reasoning traces with thinking mode on without a cap, so its thinking trace was capped at 8,192 tokens. Dataset splitting was seeded for reproducibility.

\subsection{Evaluation Metrics (RQ1, RQ2)}

On held-out data, we compare the baseline template $\pi_0$ and selected distilled template $\pi^\star$ for thinking  ($z=1$) and (especially for) more token-efficient non-thinking ($z=0$) via pass rates $\widehat{p}_{\pi,z}=\frac{\sum_{i\in\mathcal{E}} s(M(\pi(x_i),z),y_i)}{|\mathcal{E}|}$ for $\pi\in\{\pi_0,\pi^\star\}$, where $\mathcal{E}$ is a held-out test fold or release set. We abbreviate $\widehat{p}_{\pi_0,z}$ and $\widehat{p}_{\pi^\star,z}$ as $\widehat{p}_{0,z}$ and $\widehat{p}_{\star,z}$. The relevant effect is $\Delta_z=\widehat{p}_{\star,z}-\widehat{p}_{0,z}$, computed separately per mode: non-thinking gains ask whether distillation transfers some benefits of explicit reasoning into a cheaper inference setting, while thinking-mode gains ask whether the template complements an already deliberative model or saturates and introduces regressions.


We also compare distilled non-thinking inference with baseline thinking-mode
inference. For each model--task cell, we compute their pass-rate difference
and output-token savings
\(1-\frac{L_{\star,0}}{L_{0,1}}\), where \(L_{\star,0}\) and \(L_{0,1}\)
are their respective mean emitted output-token counts. These counts include
reasoning-trace and final-response tokens but exclude input-prompt tokens.
We summarize the pass-rate differences separately for Qwen and Gemma. Each
family mean gives equal weight to its eight model--task cells.

\subsection{Transfer and Common-Mode-Failure Evaluation (RQ3, RQ4)}

For RQ3, Exception, Input, and Output Prediction use five-fold cross-fitting.
In each fold, the source prompt is created from the distillation portion,
selected on the validation portion, and applied to the target model only on
that fold's held-out test examples. For Output Prediction, the target baseline
and transferred prompts use the same deterministic sample of 200 test
examples per fold. For LiveCodeBench, prompts are created and selected using
the newer v5/v6 examples, then applied unchanged to the separate v1--v4
releases. The prompt must therefore generalize both to a different model and
across benchmark releases. For each target model and release, the baseline and
transferred prompt are evaluated on exactly the same examples, so the
comparison remains paired. To save compute, we do not perform cross-model
transfer on the v5/v6 folds.

To test whether a distilled prompt transfers across models, we take a prompt
learned from a source model \(A\) and apply it to a different target model
\(B\) on the same task. We denote this transfer as \(B\leftarrow A\). All
transfer evaluations use non-thinking mode. We compare \(B\)'s pass rate with
the transferred prompt against \(B\)'s pass rate with the original task prompt
on the same examples:
\[
\Delta_{B\leftarrow A}
=\widehat{p}_{B\leftarrow A,0}-\widehat{p}_{B,0}.
\]
A positive value means that the transferred prompt helps \(B\); a negative
value means that it hurts \(B\).

Evaluations with \(A \neq B\) measure cross-model transfer and answer RQ3.
We also examine whether a model performs better with a prompt distilled from
its own behavior than with prompts transferred from other models.



A question naturally arises as to whether prompt-transfer works because of shared failures between models. The theory of 
N-version programming studies several independently developed programs for the
same task and asks whether they fail on the same inputs
\citep{ron2026version}.  When two programs fail on the same input, this is
called a \emph{common-mode failure}.  We apply the same idea to pairs of
models.
To answer RQ4 directly, we examine each directed transfer
\(B\leftarrow A\).  We first run the source model \(A\) and target model \(B\)
on exactly the same held-out examples using their baseline prompts.  Let
\(\mathcal E\) be this common set of examples. We then consider every example that target model \(B\) gets wrong.  An example
is a \emph{shared failure} if source model \(A\) is also wrong.  It is a
\emph{target-only failure} if \(A\) is correct.
A failure is \emph{rescued} if \(B\) becomes correct after receiving the
prompt distilled from \(A\).  For each directed transfer evaluation, we
calculate separate rescue rates for shared failures and target-only failures.
We then average these rates across evaluations, giving each evaluation equal
weight.

As a secondary analysis, we measure how often \(A\) and \(B\) succeed or fail
on the same examples.  For each example \(i\in\mathcal E\), we record whether
each model is correct:
\[
c_{i,A} =
\begin{cases}
1, & \text{if model \(A\) is correct on example \(i\)},\\
0, & \text{if model \(A\) is wrong on example \(i\)},
\end{cases}
\]
and define \(c_{i,B}\) in the same way for model \(B\).

We use the pairwise \(\phi\) coefficient
(from N-Version programming theory\citep{ron2026version}) to measure the similarity between the
two models' correctness patterns.  First, we compute
\[
\bar c_A=\frac{1}{|\mathcal E|}\sum_{i\in\mathcal E}c_{i,A},
\qquad
\bar c_B=\frac{1}{|\mathcal E|}\sum_{i\in\mathcal E}c_{i,B},
\]
which are the baseline pass rates of \(A\) and \(B\).  We also compute
\[
\bar c_{AB}=
\frac{1}{|\mathcal E|}
\sum_{i\in\mathcal E}c_{i,A}c_{i,B},
\]
which is the fraction of examples that both models answer correctly.  The
pairwise coefficient is
\[
\phi_{A,B} =
\frac{\bar c_{AB}-\bar c_A\bar c_B}
{\sqrt{\bar c_A(1-\bar c_A)\bar c_B(1-\bar c_B)}}.
\]

The value of \(\phi_{A,B}\) ranges from \(-1\) to \(1\).  A positive value
means that the models tend to succeed and fail on the same examples.  A value
near zero means that their outcomes have little relationship.  A negative
value means that one model tends to succeed when the other fails.  The value
is unchanged if we describe the outcomes as failures instead of successes.
It is undefined if either model is correct on every example or wrong on every
example, because that model's outcomes do not vary.

Finally, we test whether model pairs with a larger \(\phi_{A,B}\) also obtain
a larger directed transfer gain \(\Delta_{B\leftarrow A}\).  We measure this
relationship across evaluations using Spearman rank correlation.




\subsection{Additional Evaluations}

\textbf{Reasoning-budget efficiency sweep.} This comparison asks whether the distilled prompt in non-thinking mode can outperform the baseline prompt in (budgeted) thinking mode when both use roughly the same number of output tokens. For each model, task, and validation fold\footnote{We use just the validation fold to limit costs}, let \(T\) be the mean emitted tokens under the
selected distilled prompt without thinking, and let \(F\) be the mean
final-response (answer) tokens under the baseline prompt with thinking (total emitted
output minus reasoning-trace tokens). We define the medium reasoning
allowance as \(R=\max(T-F,0)\), which is the number of reasoning tokens the
(non-thinking) baseline can use while keeping its total output close to the distilled prompt.
The low, medium, and high budgets are then \(0.5R\), \(R\), and \(2R\),
rounded up to the nearest 64 native tokens. This rounding gives the reasoning
baseline at most 63 extra tokens, so the comparison is slightly conservative
for prompt distillation. We cap only the reasoning trace and keep the 14k
total generation limit, because an equal total-output cap could stop
generation before the model emits its answer. Because this sweep is expensive, we
run it only on Exception Prediction and Input Prediction using Gemma-4-E2B,
Gemma-4-E4B, and Qwen3-4B.

\textbf{Trace removal.} This ablation tests whether the reasoning trace
provides useful diagnostic information beyond the paired final answers,
correctness labels, and delta cases. We rerun Stage~1 with the same task, reference answer,
paired final responses, correctness scores, delta cases, teachers, and folds, but omit the intermediate reasoning trace and every instruction that
refers to it.  Stages~2--3 and validation selection are unchanged.  This
Full-versus-No-trace comparison covers Exception Prediction, Input Prediction,
an 800-example fixed sample of Output Prediction, and LiveCodeBench.



\textbf{GEPA baselining.}
GEPA is a recent prompt optimizer that keeps model weights fixed and improves prompts through iterative natural-language reflection and evolutionary search~\citep{agrawal2026gepa}. The original study reports that GEPA outperformed the reinforcement-learning method GRPO by 6\% on average across six tasks while requiring substantially fewer rollouts. It also surpassed MIPROv2~\citep{opsahlong2024optimizing}, the previous state-of-the-art prompt optimizer, on every benchmark and model evaluated in the GEPA study. GEPA is therefore a strong baseline for comparing Prompt Distillation with an established prompt-optimization method.

We compare the two methods using the same distillation, validation, and held-out sets across five folds, as well as the same student temperature and 14k output limit. We use GEPA~0.1.4 with non-reasoning GPT-5.4-mini as the reflector and a reflection minibatch size of three, following the original study. During optimization, the reflector receives the student's full reasoning trace and final response. Validation and held-out evaluation use the same student with reasoning disabled. We allow 1,839 student evaluations per fold, which is the smallest evaluation budget used in the GEPA study. Prompts from LiveCodeBench v5/v6 are reused in the same manner as for Prompt Distillation.

\textbf{Guidance-component removal.} This comparison tests whether the gains
come from improvement guidance, regression guards, or both. We
reuse the original splits, student baseline outputs, delta cases, Stage~1
analyses, and Stage~2 summaries.  Before Stage~3, we omit either the
improvement summary or the regression/guard summary.  We then generate three
new candidates, repeat validation selection, and evaluate the selected prompt
on the same held-out examples.  This comparison covers all four students and
five folds on Exception Prediction, Input Prediction, and LiveCodeBench.  We
report the equal-model mean change in non-thinking accuracy relative to the
complete distilled prompt.

\textbf{Local-teacher replacement.} This comparison tests whether Prompt Distillation depends on much larger frontier models as teachers. We replace both GPT teachers---GPT-5.4-mini in Stage~1 and GPT-5.4 in
Stages~2--3---with one local teacher throughout: Gemma-4-E4B, Qwen3-8B, or
Qwen3-4B.  We keep the splits, student baseline outputs, and delta cases fixed,
but rerun all teacher-dependent stages, validation selection, and held-out
evaluation.  This comparison covers all four students and five folds on
Exception Prediction and Input Prediction.  Reported deficits are equal-model
mean differences from the GPT-teacher condition.  We compare paired held-out
outcomes with two-sided McNemar tests.

\section{Results}
\label{sec:results}

\subsection{RQ1--RQ2: Effectiveness and inference efficiency}

\begin{table}[tb]
\centering
\footnotesize
\setlength{\tabcolsep}{2pt}
\begin{tabular}{@{}lcccc@{}}
\toprule
Task family & Qwen3-4B & Qwen3-8B & Gemma-4-E4B & Gemma-4-E2B \\
\midrule
LiveCodeBench$^\dagger$ & $0.396\!\rightarrow\!0.419$ (+2.3) & $0.426\!\rightarrow\!0.459$ (+3.3)$^{**}$ & $0.637\!\rightarrow\!0.685$ (+4.8)$^{***}$ & $0.525\!\rightarrow\!0.569$ (+4.4)$^{***}$ \\
\quad\emph{output-token savings} & 86.5\% & 83.4\% & 39.8\% & 44.8\% \\
Exception Prediction & $0.447\!\rightarrow\!0.562$ (+11.5)$^{***}$ & $0.565\!\rightarrow\!0.639$ (+7.4)$^{***}$ & $0.715\!\rightarrow\!0.739$ (+2.4)$^*$ & $0.618\!\rightarrow\!0.645$ (+2.8)$^*$ \\
\quad\emph{output-token savings} & 71.7\% & 75.1\% & 40.6\% & 32.1\% \\
Input Prediction & $0.596\!\rightarrow\!0.616$ (+2.0) & $0.569\!\rightarrow\!0.648$ (+7.9)$^{***}$ & $0.890\!\rightarrow\!0.905$ (+1.5) & $0.608\!\rightarrow\!0.635$ (+2.8) \\
\quad\emph{output-token savings} & 77.4\% & 71.3\% & 29.6\% & 35.7\% \\
Output Prediction & $0.720\!\rightarrow\!0.719$ (-0.0) & $0.746\!\rightarrow\!0.759$ (+1.3)$^{**}$ & $0.898\!\rightarrow\!0.898$ (+0.0) & $0.816\!\rightarrow\!0.834$ (+1.8)$^{***}$ \\
\quad\emph{output-token savings} & 71.2\% & 77.0\% & 35.8\% & 29.3\% \\
\bottomrule
\end{tabular}
\caption{
RQ1 and RQ2 held-out results. Per task family, the first row reports non-thinking baseline $\rightarrow$ distilled pass rate with the signed percentage-point difference (computed before rounding; nominal paired exact McNemar tests: $^*$ $p<0.05$, $^{**}$ $p<0.01$, $^{***}$ $p<0.001$). The second row reports the output-token savings of distilled non-thinking inference relative to baseline thinking-mode inference (maximum thinking budget). $^\dagger$Pools the v5/v6 fold tests with the v1--v4 release evaluation.}
\label{tab:rq1_effects}
\end{table}

Prompt distillation generally helps; across four task families and four student models, 15 of 16 non-thinking model--task comparisons show performance gains, with a mean gain of 3.5 pass-rate points; 10 are significant under paired exact McNemar tests (Table~\ref{tab:rq1_effects}). Gains are largest on Exception Prediction (up to 11.5 points for Qwen3-4B), and the pooled LiveCodeBench estimates, which include the independent v1--v4 release evaluation, are positive for every model (+2.3 to +4.8). Effect signs remain positive under every leave-one-fold/release-out omission for 12 of the 15 positive effects (that is, the gain remains positive no matter which single test fold is omitted). Output Prediction, with comparatively high baselines, averages only +0.8 points.

With thinking mode, benefits are smaller and less consistent: 12 of 16 comparisons see an improvement with the distilled prompt (mean +1.6 points). Exception Prediction and Input Prediction remain positive for every model, while LiveCodeBench and Output Prediction include negative estimates.

For RQ2, Table~\ref{tab:rq1_effects} shows substantial output-token savings in
every model--task comparison. Distilled non-thinking inference uses 56.3\%
fewer output tokens than maximum-budget thinking on average. The savings range
from 29.3\% to 86.5\%, and are largest on LiveCodeBench at the task level
(63.6\% on average). The family-level savings are significant (for both Qwen
(\(p=.0044\)) and Gemma (\(p=.0147\))).

The accuracy cost differs sharply between the two model families. Qwen saves
76.7\% of its output tokens on average, but maximum-budget thinking is 16.8
pass-rate points more accurate and wins all eight comparisons. Gemma saves
36.0\% of its output tokens while giving up only 1.3 points on average (relative to full thinking);
full thinking wins six of eight comparisons. This difference
suggests that the on/off thinking comparison may give Qwen much more
reasoning computation than needed. We therefore next test smaller reasoning
budgets and compare their accuracy and token use with distilled non-thinking
inference.

\begin{figure}[tb]
\centering
\includegraphics[width=0.96\linewidth]{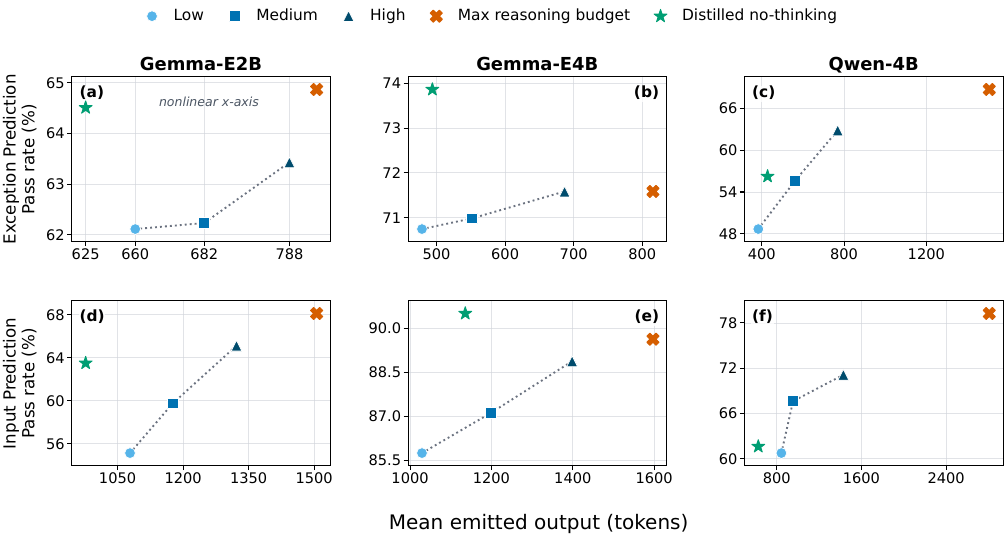}
\caption{Reasoning-budget efficiency on held-out Exception
and Input Prediction examples. Low, medium, and high cap only the reasoning
trace; the horizontal axis reports the observed reasoning-plus-final-response
output. Stars denote distilled non-thinking inference and crosses denote the
maximum-budget thinking condition. All conditions retain the 14k total
generation limit. Only Qwen has a separate maximum reasoning-trace budget:
8,192 native tokens, a liberal limit that is rarely reached. Gemma has no
separate trace limit. Points pool five held-out folds. Panel (a) uses the
explicitly labeled nonlinear horizontal scale. Qwen3-8B is not included in
this controlled sweep.}
\Description{Six panels compare pass rate against mean emitted output tokens
for three student models on Exception Prediction and Input Prediction.
Increasing the reasoning budget generally raises accuracy and token use.
Distilled non-thinking inference is shown with green stars, and
maximum-budget thinking with orange crosses. All runs have a 14k total
generation limit; only Qwen has a separate 8,192-token reasoning-trace cap.}
\label{fig:reasoning-budget-efficiency}
\end{figure}

\textbf{Reasoning-budget sensitivity.}
Figure~\ref{fig:reasoning-budget-efficiency} asks whether reducing the
reasoning budget can match the efficiency of prompt distillation. Increasing
the reasoning budget raises both accuracy and output length in all six
model--task cells. The medium budget is chosen to approximate the output
length of distilled non-thinking inference. However, it limits only the text
inside the thinking trace, not the total response. Some models continue
step-by-step reasoning after the closing \texttt{</think>} tag. As a result,
medium-budget, and sometimes low-budget, reasoning can still produce longer
outputs than distilled non-thinking inference.

Prompt distillation provides a better accuracy--token tradeoff. Compared with
medium-budget reasoning, it emits 57--329 fewer output tokens in all six
cells. It is also more accurate in five of the six cells. Thus, in five cells,
prompt distillation is both shorter and more accurate.

The only exception is Qwen3-4B on Input Prediction. Medium-budget reasoning is
6.0 pass-rate points more accurate, but it emits 329 more tokens on average.
The distilled prompt also emits 218 fewer tokens than even the low-budget
reasoning condition, while being 0.9 points more accurate. The additional
accuracy from the medium budget therefore comes at a substantial token cost.
Across all six cells, prompt distillation provides an attractive performance tradeoff.

\begin{figure}[tb]
\centering
\includegraphics[width=0.98\linewidth]{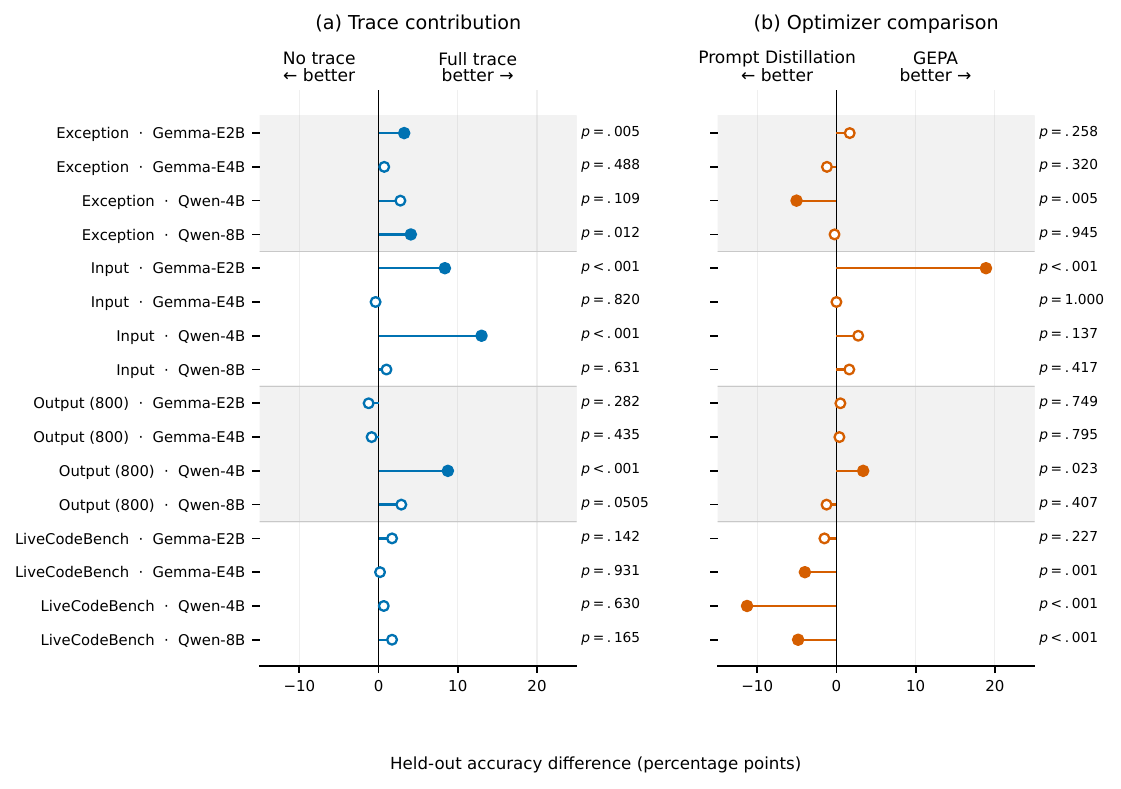}
\caption{Controlled comparisons of paired held-out effects.
(a) Full prompt distillation versus the No-trace condition.
(b) GEPA versus Prompt Distillation. Lines connect zero to each accuracy
difference, and labels report unadjusted two-sided exact McNemar
\(p\)-values. Filled points denote \(p<0.05\).
Output Prediction uses the fixed 800-example sample, and LiveCodeBench pools
v5/v6 with v1--v4.}
\Description{Two effect-size dot plots with unadjusted exact McNemar p-values.
(a) compares full prompt distillation against an otherwise identical
No-trace condition. (b) compares GEPA against Prompt Distillation.
Effects vary substantially across tasks and models.}
\label{fig:controlled-comparisons}
\end{figure}

\subsection{Does the reasoning trace matter?}
We next test whether the intermediate reasoning trace adds useful information beyond the paired final answers and their correctness labels.
Figure~\ref{fig:controlled-comparisons}(a) reports the trace-removal
comparison.
Full traces outperform No trace in 13 of 16 task--model cells, with an
equal-cell mean of 2.9 pass-rate points.  The largest effects are Input/Qwen-4B
(+13.0), sampled Output/Qwen-4B (+8.75), and Input/Gemma-E2B (+8.38).
Two-sided exact McNemar tests give \(p<0.05\) for 5 of the 16
task--model cells: Exception/Gemma-E2B, Exception/Qwen-8B,
Input/Gemma-E2B, Input/Qwen-4B, and sampled Output/Qwen-4B. Thus traces provide
useful information on average, but their value is heterogeneous.


\subsection{Comparison with GEPA}

\paragraph{Performance on benchmarks.}
Figure~\ref{fig:controlled-comparisons}(b) compares GEPA with Prompt
Distillation. GEPA is more accurate in 7 of 16 task--model cells, ties in 1,
and is less accurate in 8. Averaged across the four tasks, the mean
GEPA-minus--Prompt Distillation difference is less than 0.01 pass-rate
points. Two-sided exact
McNemar tests find significant differences in six cells.
GEPA is better for Input Prediction with Gemma-E2B (+18.9 points) and sampled
Output Prediction with Qwen-4B (+3.4). Prompt Distillation is better for
Exception Prediction with Qwen-4B (+5.0) and for LiveCodeBench with Gemma-E4B
(+4.0), Qwen-4B (+11.3), and Qwen-8B (+4.8). Thus, the methods perform
similarly on average, but Prompt Distillation is clearly stronger on most
LiveCodeBench comparisons.

\paragraph{Cost comparison.}
We first compare prompt size across 80 selected prompts: four students, five
folds, and four tasks.
Prompt Distillation adds \(55.9\pm11.2\) tokens to the original task prompt
(mean\(\pm\)SD), while GEPA adds \(701.2\pm545.8\).  Thus GEPA's selected
instructions are about an order of magnitude longer and add more input tokens
to every later request.

For one-time optimization cost, the student workloads are similar: Prompt Distillation makes 122,072 student
calls and GEPA makes 124,468.  We reconstruct teacher input and output token costs from the recorded token counts, using the prices of the models used in the experiments (GPT-5.4-mini: $0.75/$4.50 per million input/output tokens; GPT-5.4: $2.50/$15.00 per million input/output tokens). Prompt Distillation costs \$41.93 in
total, compared with \$53.83 for GEPA---we are 22.1\% less.  Prompt Distillation is
cheaper on three tasks; GEPA is cheaper only on Input Prediction (due to high number of delta cases in Prompt Distillation).

For deployment cost, optimization is paid once, but the selected prompt is sent with every
request.  We therefore count the complete non-thinking input and output for
each held-out example.  Prompt Distillation inputs come directly from the
stored predictions.  GEPA inputs use the same examples with the selected GEPA
instruction substituted for the baseline instruction.  We then pool the five
folds and sum the four tasks within each two-model family.

\begin{figure}[tb]
\centering
\includegraphics[width=0.65\linewidth]{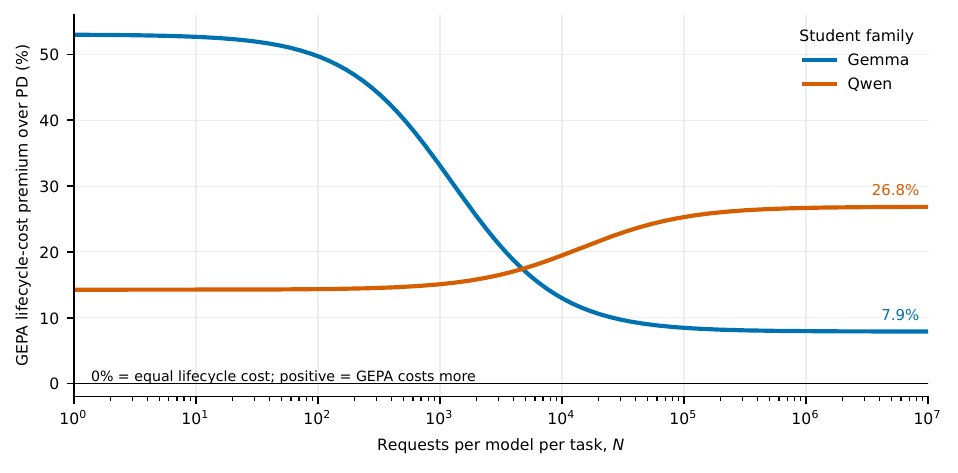}
\caption{Combined optimization and deployment cost over all four tasks.
\(N\) requests per model per task correspond to \(8N\) calls per family.
GEPA is generally more expensive than our Prompt Distillation approach (PD), as
indicated by values remaining above zero.}
\Description{A log-scale plot from one to ten million requests per model per
task.  GEPA's cost premium remains positive for both model families: the
Gemma curve approaches about 8 percent and the Qwen curve about 27 percent.}
\label{fig:lifecycle-cost-scaling}
\end{figure}

Let \(N\) be the number of requests to each model on each task.  Each family
therefore represents \(8N\) calls: two models, four tasks, and \(N\) requests.
The calculation is simply
\[
\text{total cost}(N)=\text{one-time teacher cost}
  +N\times\text{inference cost}.
\]
Figure~\ref{fig:lifecycle-cost-scaling} shows GEPA's percentage premium over
Prompt Distillation.  Zero means equal cost; positive values mean GEPA costs
more.

Together AI is a cloud inference platform that hosts open-weight
models.\footnote{Together AI serverless prices per million tokens,
\url{https://www.together.ai/pricing}.} It does not list prices for our exact
student models, and we did not find these prices on other cloud platforms.
We therefore use larger models from the same families as pricing proxies.
We use Gemma-4-31B at \$0.39/\$0.97 per million input/output tokens and
Qwen3.5-9B at \$0.17/\$0.25.  These are not bills for our local serving; they
are a sensitivity analysis using plausible upper-bound prices.

This choice is conservative for Prompt Distillation.  GEPA uses many more
input tokens, but fewer output tokens.  Because the proxy prices charge more
for output than input, they give GEPA extra credit for those shorter outputs.
Even with that favorable weighting, GEPA has both a higher one-time cost and
a higher per-request cost in both families.  The curves therefore cannot
cross: GEPA remains more expensive for every \(N\geq0\).  If the smaller
students have lower output/input price ratios, Prompt Distillation's advantage
would be larger.  We do not assume that price ratio always grows with model
size.

Teacher cost also depends on the number of delta cases. Prompt Distillation sends every improvement and regression case to its
Stage~1 teacher, so its teacher cost increases with the number of delta cases.
For each task, we use the 20 student--fold runs to fit a simple linear
relationship between the number of delta cases and the teacher-cost difference
between GEPA and Prompt Distillation. The fitted costs become equal at about
125 cases for Exception Prediction, 64 for Input Prediction, 134 for sampled
Output Prediction, and 221 for LiveCodeBench. Below these values, Prompt
Distillation is estimated to be cheaper. All observed LiveCodeBench folds have
only 39--63 cases and favor Prompt Distillation; its 221-case cutoff lies
outside the observed range. These fitted values describe our runs and are not
general guarantees.

More delta cases therefore have a clear cost. However, that added cost would
be worthwhile if more cases produced a larger accuracy gain. We test this by
checking whether folds with more delta cases achieve larger held-out gains.

\begingroup
\setlength{\intextsep}{3pt}
\setlength{\columnsep}{8pt}
\begin{wrapfigure}{r}{0.50\linewidth}
\vspace{-0.8\baselineskip}
\centering
\includegraphics[width=\linewidth]{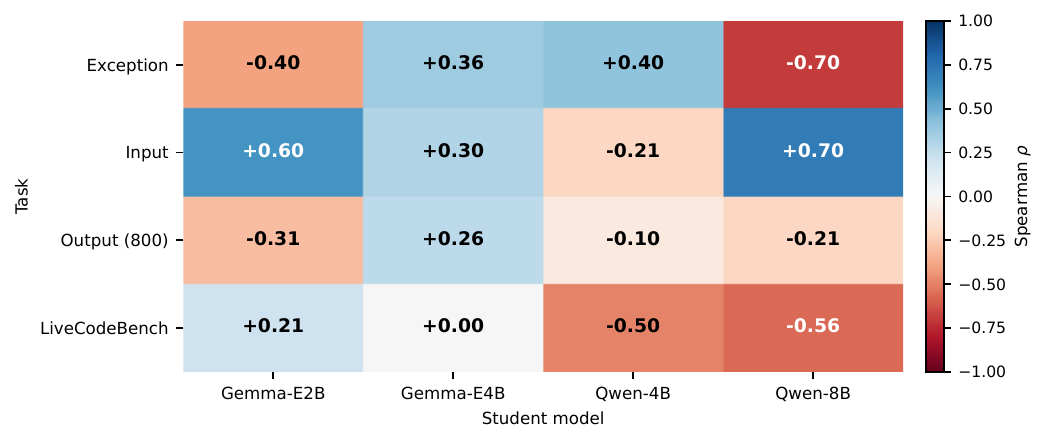}
\caption{Fold-level association between delta-case count and held-out Prompt
Distillation gain. Each cell reports Spearman's \(\rho\) over five folds; the
small sample makes this an exploratory diagnostic.}
\Description{A four-by-four heatmap of correlations between delta-case count
and held-out gain. Values have mixed signs and range from minus 0.70 to plus
0.70.}
\label{fig:delta-gain-correlations}
\vspace{-0.6\baselineskip}
\end{wrapfigure}
The relationship is not consistent. Across task--student cells, the five-fold correlations
between case count and held-out gain range from \(-0.70\) to \(+0.70\)
(Figure~\ref{fig:delta-gain-correlations}).  After pooling students within
each task, the correlations are 0.73 for Exception, 0.39 for Input, 0.12 for
sampled Output, and \(-0.07\) for LiveCodeBench.  More delta cases therefore
increase teacher cost, but do not reliably predict a larger accuracy gain.
\subsection{Component and teacher ablations.}
Removing improvement guidance lowers mean non-thinking accuracy by 3.18 points
on Exception Prediction, 4.53 on Input Prediction, and 2.78 on LiveCodeBench.
Regression guidance is task-dependent: removing it lowers Exception and
LiveCodeBench by 1.41 and 1.54 points, but improves Input by 3.59 points.
Replacing the GPT teachers with Gemma-4-E4B, Qwen3-8B, or Qwen3-4B also lowers
average non-thinking accuracy: deficits versus GPT range from 4.47--6.56
points on Exception and 3.81--11.41 on Input.  These teacher comparisons cover
only two tasks.
\par
\endgroup

\subsection{RQ3: Cross-model transfer}
\begin{table}[tb]
\centering
\small
\setlength{\tabcolsep}{3pt}
\begin{tabular}{@{}lrrrrrrr@{}}
\toprule
Task & Transfer evals. & Pos./Neg. & Base & Transfer & Mean $\Delta$ $\downarrow$ & Median $\Delta$ & Wilcoxon $p$ \\
\midrule
Exception pred. & 60 & 48/9 & 58.6 & 63.0 & +4.4 & +3.6 & $<.001$ \\
LiveCodeBench$^\dagger$ & 48 & 34/9 & 53.5 & 56.1 & +2.7 & +2.9 & $<.001$ \\
Output pred. & 60 & 34/25 & 80.1 & 79.4 & -0.7 & +1.0 & .071 \\
Input pred. & 60 & 32/25 & 66.6 & 65.8 & -0.7 & +0.6 & .207 \\
\bottomrule
\end{tabular}
\caption{RQ3 non-thinking cross-model transfer. \(N\) is the number of
directed evaluations; \(+/-\) counts positive/negative effects, with remaining
evaluations tied. Rates are percentages and deltas are percentage points.
\(p\) is the one-sided Wilcoxon signed-rank test for positive transfer.
\(^{\dagger}\)Evaluation on unseen releases.}
\label{tab:rq3_transfer}
\end{table}

With four models, each held-out fold or release has 4 target models and 3
non-self source models, giving 12 directed evaluations. LiveCodeBench uses
four unseen releases, giving 48 evaluations. Each other task uses five folds,
giving 60 evaluations. Output Prediction uses 200 held-out examples per fold.
For RQ3, cross-model transfer is positive on two of four task families
(Table~\ref{tab:rq3_transfer}). Exception Prediction is strongest: 48 of 60
effects are positive, with a mean benefit of 4.4 points. LiveCodeBench gains
2.7 points on average. Input and Output Prediction have slightly negative
means but positive medians, with similar numbers of positive and negative
effects.

On the same held-out folds or releases, a model's own distilled
prompt has a larger mean gain in every task. The own-prompt versus cross-model
gains are 6.0 versus 4.4 points for Exception Prediction, 3.6 versus 2.7 for
LiveCodeBench, 3.5 versus \(-0.7\) for Input Prediction, and 0.8 versus
\(-0.7\) for Output Prediction. Across all evaluations, the mean gain is 3.5
points with the model's own prompt and 1.3 points with cross-model prompts.

\subsection{RQ4: Common-mode failures and transfer}
\begingroup
\setlength{\intextsep}{3pt}
\setlength{\columnsep}{8pt}
\begin{wraptable}{r}{0.58\linewidth}
\vspace{-0.8\baselineskip}
\centering
\footnotesize
\setlength{\tabcolsep}{2pt}
\begin{tabular}{@{}lrrr@{}}
\toprule
Scope & Mean $\phi$ & $\rho_\phi$ $\downarrow$ & \(p\) \\
\midrule
LiveCodeBench$^\dagger$ & 0.52 & 0.22 & .124 \\
Output pred. & 0.42 & 0.17 & .185 \\
Input pred. & 0.29 & 0.10 & .428 \\
Exception pred. & 0.44 & -0.18 & .172 \\
\midrule
All evaluations & 0.41 & 0.10 & .131 \\
\bottomrule
\end{tabular}
\caption{Secondary RQ4 analysis: Spearman association between pairwise
failure \(\phi\) and directed transfer gain \(\Delta_{B\leftarrow A}\).
Two-sided \(p\)-values test \(H_0:\rho_\phi=0\).
\(^{\dagger}\)LiveCodeBench release evaluations.}
\label{tab:rq4_common_failure_transfer}
\vspace{-0.6\baselineskip}
\end{wraptable}

Does $B \leftarrow A$
prompt transfer more effectively rescue examples where both $A$ and $B$ fail? In fact,
transferred prompts only rescue a mean of 21.3\% of shared failures, compared with
47.3\% of target-only failures across the 228 directed evaluations.  The
shared-failure rescue rate is lower in every task family: 13.1\% versus 48.6\%
on LiveCodeBench, 17.6\% versus 40.0\% on Exception Prediction, 30.1\% versus
52.3\% on Input Prediction, and 22.7\% versus 48.6\% on Output Prediction.
Thus, the results do not support shared failure as the main reason that
transferred prompts help.

As a secondary analysis, the failure-correlation measure \(\phi\) ranges from 0.29 on Input Prediction to 0.52 on LiveCodeBench (Table~\ref{tab:rq4_common_failure_transfer}). This indicates substantial common-mode failure between models: models fail surprisingly often on the same inputs. Its relationship with transfer gain is weak: the descriptive aggregate over 228 directed transfer evaluations has a rank correlation of 0.10, and per-task correlations are mixed. We thus fail to observe a strong relationship between common-mode failure intensity and prompt-transfer effectiveness. 
\par
\endgroup

\section{Discussion}
\label{sec:discussion}

Overall, our findings suggest that advisory prompts can capture some benefits of thinking behavior at lower cost. The pipeline distills paired thinking/non-thinking behavior changes into helpful prompts; however, distilled prompts do not always yield the full gains of costlier thinking mode (\autoref{fig:reasoning-budget-efficiency}). 
The cross-model transfer analyses of RQ3 and RQ4 suggest that prompts only
sometimes beneficially transfer, and  not always as predicted by common-mode failure.   

\paragraph{Distilled prompts act as compact procedural guidance.}
Advisory prompts help most in \emph{non-thinking} mode:
15 of 16 comparisons improve.
Thinking-mode effects are smaller and sometimes negative, partly because the
stronger thinking-mode baselines leave less room for improvement. In an
exploratory analysis of 96 model-specific
held-out sets, where each set is one model evaluated on one test fold or
LiveCodeBench release, the gain due to thinking \emph{per se} has a positive
Spearman correlation with the non-thinking distillation gain
\(\Delta_0\) (\(\rho=0.35\),
\(p<0.001\)). The content of the prompts fits this picture and matches the
design intent: from each delta case, the teacher abstracts what the model
should do or avoid, and the resulting prompts are short, generic instructions
averaging an increase of
\(55.9\pm11.2\) prompt-template tokens
(mean\(\pm\)SD), with no
problem-specific reasoning. In a keyword audit of the
80 selected prompts,
56\% ask the model to track state or
transitions, 48\% to work through concrete
examples or dry runs, and 31\% to check
edge cases. Guiding the model to do these things more cheaply recovers part of the benefit of thinking mode, echoing evidence from code generation that concise, structured plans retain most of the accuracy of reasoning chains roughly ten times longer \citep{jin2025understanding}. Unlike those per-problem plans, our instructions are fixed for the whole task. As we discuss
below, this ``procedural guidance'' also  helps with transferability.

\paragraph{Distillation defines an accuracy--cost operating point.}
The RQ2 results in Figure~\ref{fig:reasoning-budget-efficiency} show that thinking/non-thinking
is a coarse distinction.
The optimization-side results add a practical qualification:
because Prompt Distillation teacher cost scales with the number of delta
cases, capping each fold at 64 total improvement-plus-regression cases keeps it
below every fitted task-specific GEPA parity point in our data.
Thinking mode traces are often far longer than a problem requires \citep{chen2024think,sui2025stop}, and the accuracy--cost tradeoff differs by model family. Table~\ref{tab:rq1_effects} reports the output-token savings. These savings average 76.7\% for the Qwen3 students and 36.0\% for the Gemma-4 students. Compared with baseline thinking mode, distilled non-thinking inference gives up 16.8 pass-rate points on average for Qwen3, but only 1.3 points for Gemma-4. This split is consistent with how differently the families engineer thinking: Qwen3 trains it through a dedicated multi-stage pipeline and exposes an explicit thinking budget to manage its cost \citep{qwen2025qwen3}, whereas Gemma-4 adds thinking on top of standard post-training without any described length control \citep{gemmateam2026gemma}. The savings do not come from just shortening the outputs: in
86 of 96 model-specific held-out
sets, where each set is one model evaluated on one fold or LiveCodeBench
release, distilled outputs are in fact longer than the non-thinking baseline outputs. Hybrid models are known to separate the two modes only imperfectly: even with thinking disabled, reasoning steps can appear in the visible response, outside the \texttt{<think>} delimiters \citep{wang2025demystifying}. The extra length is consistent with this: the prompt appears to elicit brief working directly in the answer, a far more concise substitute for the full trace. Distilled (non-thinking) prompts  thus offer a cheaper default, while
escalating harder or higher-stakes cases to high-budgeted thinking mode.

\paragraph{Why might GEPA struggle on heterogeneous coding tasks?}
We examine one LiveCodeBench example to illustrate a possible reason.  The
example uses Qwen3-4B. The problem, ``Find the Original Typed String I,'' appears
only in the final test set.

The problem asks the model to count possible intended strings after at most
one accidental long keypress.  For example, the observed string
\texttt{aaaa} has four possible intended strings:
\texttt{aaaa}, \texttt{aaa}, \texttt{aa}, and \texttt{a}.  The base-prompt
response fails this problem.  The GEPA response also fails.  The Prompt
Distillation response passes all 3 public and 31 private tests.  The complete
Prompt Distillation template has 81 tokens, while the GEPA template has 918.

The GEPA instruction includes detailed solutions to three unrelated problems:
"array division", "weighted adjacent swaps", and "robot-grid movement".  It also
gives this general rule: ``when a problem involves `at most \(K\) special
actions,' model \(K\) explicitly in the DP state.''  Here, DP means dynamic
programming, a method that stores partial results in a state table.  Because
the held-out problem says ``at most once,'' the generated response follows
this DP advice:
\begin{lstlisting}[language=Python,basicstyle=\scriptsize\ttfamily,numbers=none,xleftmargin=1em]
# GEPA response
dp[i] = dp[i - 1]
if word[i - 1] == word[i - 2]:
    dp[i] += dp[i - 2]
return dp[n]
\end{lstlisting}
This state does not record whether the one allowed mistake has already been
used.  It therefore combines choices from several repeated-character
positions, as if more than one mistake were allowed.  It returns 5 for
\texttt{aaaa}, instead of 4, and 10 for \texttt{abbcccc}, instead of 5.

The Prompt Distillation instruction is shorter and does not name earlier
problems.  It asks the model to derive the exact counting rule, check the rule
against the problem statement, and test a difficult example.  The response
rejects dynamic programming and instead counts repeated characters:
\begin{lstlisting}[language=Python,basicstyle=\scriptsize\ttfamily,numbers=none,xleftmargin=1em]
# Prompt Distillation response
total += run_length - 1
return total + 1  # include the unchanged original
\end{lstlisting}
There is always one possibility in which the observed string is unchanged.
A repeated run of length \(L\) adds \(L-1\) more possibilities, obtained by
shortening that one run.  The resulting formula is
\(1+\sum_{\text{runs}}(\text{run length}-1)\).  It gives 4 for
\texttt{aaaa}.  For \texttt{abbcccc}, it gives
\(1+(2-1)+(4-1)=5\).

This single example does not prove why GEPA fails.  Each condition uses only
one stochastic generation.  However, the example suggests that GEPA can carry
detailed recipes from earlier problems into a new and different problem.
Those recipes may point the model toward the wrong method.  Prompt Distillation
instead combines many observations into shorter, task-level advice.  Future
work could test this explanation by removing problem-specific sections from
GEPA prompts or by comparing prompts of similar length.

\paragraph{Transfer varies by task.}
The positive RQ3 (model-transfer) results on Exception Prediction and LiveCodeBench are consistent with
\emph{procedural guidance} being reusable across the evaluated models. We hypothesize that procedural guidance specifying checks with discrete semantics, such as operator compatibility, API existence, and interface preservation, is easier to reuse across models than instructions that describe a model-specific strategy. Therefore, procedural guidance prompts also help on LiveCodeBench, a code generation task. This distilled prompt provides procedural guidance to include steps such as checking invariants and verifying output formats against the samples.
By contrast, on the Input and Output Prediction tasks, neither thinking
nor non-thinking were consistently better across models, so the teacher models struggled to produce consistent 
distilled procedural
guidance that consistently improved performance across models.  

\paragraph{Shared failure is not sufficient to explain transfer.}
The  weak (RQ4) association of common-mode failure with prompt transfer rescue has a structural component: failure $\phi$ is symmetric, whereas prompt transfer is directional. Two models can fail in the same examples, but for different reasons, and react differently to the same transferred prompt. Across the four tasks, the Gemma-4-E4B and Gemma-4-E2B pair has a mean pairwise $\phi$ of 0.45, yet E4B-to-E2B transfer gains 4.2 points on average, while the reverse direction loses 2.4 points. The RQ4 results also show that shared failures are less likely to be rescued than target-only failures. Classical N-version programming theory explains this: when difficulty varies across inputs, independently developed versions can both fail on intrinsically hard cases, even without any shared faulty designs \citep{eckhardt1985theoretical}. From this viewpoint, positive $\phi$ may just indicate a set of shared hard examples, and target-only failures are just the cases which the source model solves, so its prompt is more likely to encode the relevant remedy. $\phi$ thus tells us where two models fail together, and hence where a transferred prompt could help, but not whether the source prompt encodes the right remedy or whether the target will follow it; direction-specific predictors such as prompt-content compatibility are left to future study.

\section{Related Work}
\label{sec:related_work}
\paragraph{Prompt optimization from failures.}

Automatic prompt optimization changes a model's instructions while keeping
its weights fixed. Some methods generate several instructions and select the
best one \citep{zhou2023large}. Others verbally describe  errors in a sample and
use this feedback as a textual ``gradient'' for revising the prompt
\citep{pryzant2023automatic}. GEPA repeatedly reflects on model runs and
evolves the prompt \citep{agrawal2026gepa}. This approach can rival
reinforcement-learning post-training and has also been used as an
inference-time search method to improve generated programs on
hardware-specific code-generation tasks (kernels)~\citep{agrawal2026gepa}.
\citet{ramnath2025systematic} survey this broader area.
Several methods learn reusable guidance from model mistakes. They reflect on
earlier model trials \citep{shinn2023reflexion}, extract principles from errors
\citep{zhang2024context}, or summarize experience from previous agent runs
\citep{zhao2024expel}. However, a model still cannot reliably correct itself
using only its own feedback, especially on code
\citep{huang2024large,olausson2024self}.
ContraPrompt is most closely related to our method. It compares successful
and failed reasoning traces for the same input, then turns their differences
into prompt rules \citep{rishav2026contraprompt}. Other recent methods also
learn from contrasting examples or from memories of earlier attempts
\citep{li2026learning,koh2026contrastive,nasvytis2026core}.

Our method differs in two main ways. First, we compare the thinking and
non-thinking responses of the same model. We use both improvements and
regressions. If thinking changes a correct answer into an incorrect one, we
turn that failure into a guard instruction. Second, we study not only
accuracy, but also inference cost, cross-model transfer, and common-mode
failures.

\paragraph{Thinking mode and efficient reasoning.}

Explicit reasoning improves accuracy \citep{wei2022chain,deepseekai2025deepseek}, and hybrid reasoning models expose a togglable thinking mode \citep{qwen2025qwen3,anthropic2025claude}. But thinking helps unevenly, on code in particular \citep{sprague2025cot,jin2025understanding}, and can even hurt \citep{li2025thinking,gema2025inverse}; precisely why we 
consider both improvement and regression cases. Remedies for overthinking \citep{sui2025stop} include budget forcing \citep{muennighoff2025s1}, prompts optimized for brevity \citep{yu2025premise}, and in-context reuse of distilled ``behaviors'' \citep{didolkar2025metacognitive}, while mode switching itself remains imperfect \citep{wang2025demystifying}. These compress reasoning inside the model or its outputs; we instead distill correctness deltas between a model's own inference modes into a fixed advisory prompt without weight or gradient access.

\paragraph{Reasoning distillation.}
Classic distillation transfers reasoning by training on rationales \citep{hsieh2023distilling}; see \citet{xu2024survey} for a survey. Concurrent non-parametric variants encode teacher reasoning patterns as system-prompt instructions \citep{badhe2026prompt} or distill task data into prompts \citep{dyagin2025automatic}. Unlike these, we distill the behavioral delta between a student's own inference modes, with no weight updates anywhere; to our knowledge, prompt distillation of either kind has not previously been evaluated on execution-checked software-engineering tasks.

\paragraph{Prompt transfer and common-mode failures.}
Prompt effects are large and weakly correlated across models \citep{sclar2024quantifying,mizrahi2024state}, yet discrete prompts sometimes transfer \citep{hong2024dp,melamed2024prompts} even as machine-optimized prompts often degrade on new models \citep{rakotonirina2023discrete,wang2025promptbridge}; for jailbreak prompts, transferability is predicted by representation similarity between models \citep{angell2026jailbreak}. We contribute a transfer study for failure-distilled instructions, testing a behavioral analogue of such predictors: shared failures. Our analysis borrows the common-mode failure lens of N-version programming \citep{chen1978version}, in which independently developed versions fail together far more than independence predicts \citep{knight1986experimental,eckhardt1985theoretical}; modern analogues find correlated errors across LLM-based code generators \citep{ron2026version,goel2025great,nogueira2026systematic}. We repurpose this lens diagnostically: our models show positive pairwise failure association, yet co-failure only weakly predicts cross-model prompt-transfer gain.

\section{Threats to Validity}

\paragraph{Construct validity.}
Our main outcome is pass rate. This captures functional correctness, but it
does not measure other software engineering qualities such as maintainability,
security, or real-world usability. Our efficiency comparisons also focus on
token counts and token-based pricing proxies rather than wall-clock latency,
memory use, or energy consumption.

\paragraph{Internal and statistical validity.}
Each condition uses a single stochastic generation at temperature 0.7, so
some observed differences may reflect sampling variability. The
teacher-generated diagnoses are post-hoc natural-language explanations and
were not independently validated. The trace-removal ablation isolates access
to the intermediate trace, but it does not prove that any specific teacher
explanation is causally correct.

\paragraph{External validity.}
Our experiments evaluate only four relatively small student models, two
teacher models, and four predominantly Python-oriented software engineering
tasks. The results may therefore not generalize to larger models, other
programming languages, interactive software engineering workflows, or
different decoding settings. Cross-model transfer is often helpful but
directional, and it can sometimes reduce performance, so distilled prompts
should be validated for each target model before deployment. The
reasoning-budget sweep also covers only two tasks, three students, and three
validation-derived budget levels, so it does not establish a full
accuracy--compute frontier across all evaluated settings. Our GEPA comparison
also uses one matched configuration and one evaluation budget, so it should
be read as a controlled baseline comparison rather than an exhaustive search
over GEPA settings.

\section{Conclusion}

We introduced a purely inference-time pipeline that distills thinking/non-thinking correctness deltas into short advisory prompts: compact procedural guidance rather than reproduced reasoning. On held-out data, 
\emph{non-thinking gains appear in 15 of 16 model--task comparisons} (mean +3.5 points) across four student models and four software-engineering task families. Relative to baseline full thinking, distilled non-thinking inference saves 56.3\% of output tokens on average. The average accuracy cost is 16.8 points for the Qwen3 students and 1.3 points for Gemma-4. Cross-model transfer is often positive but directional; pairwise failure $\phi$ is only weakly associated with transfer gains, and shared failures are rescued less often than target-only failures. These findings support prompt distillation as a weight-update-free, task-specific intervention. Future work should add mode-specific selection with a baseline fallback, seek direction-specific transfer predictors, and pair distilled prompts with execution feedback for shared failures. GEPA and
Prompt Distillation each win in different settings; GEPA uses much longer
optimized instructions and has higher aggregate teacher cost under matched
student-evaluation workloads.  Prompt Distillation is therefore best viewed as
one useful, auditable operating point, not a universal replacement for thinking or general prompt optimization.

\section*{Data Availability}
The anonymous replication package is available at
\url{https://doi.org/10.5281/zenodo.23117128}. It includes code, retained
artifacts, manifests, and scripts sufficient to audit and reproduce the reported
results, given required model and teacher API access. Full datasets are
retrieved from their canonical public sources.


\bibliographystyle{ACM-Reference-Format}
\bibliography{references}

\end{document}